\documentclass[twocolumn]{aastex63}     

\usepackage{epsfig}
\usepackage{graphicx}
\usepackage{epstopdf}

\usepackage{amsmath}
\usepackage{amssymb}

\usepackage{comment}
\usepackage{natbib}

\usepackage{color}

\newcommand\beq{\begin{equation}}
\newcommand\eeq{\end{equation}}

\newcommand{\ben}{\begin{eqnarray}}
\newcommand{\een}{\end{eqnarray}}
\newcommand{\benn}{\begin{eqnarray*}}
\newcommand{\eenn}{\end{eqnarray*}}

\allowdisplaybreaks

\shorttitle{Imbalanced collisionless Alfv\ 'en wave turbulence}
\shortauthors{Miloshevich et al.}

\begin{document}

\title{Modeling imbalanced collisionless Alfv\'en wave turbulence with nonlinear diffusion equations}

\author{G. Miloshevich}
\affiliation{Universit\'e C\^ote d'Azur, CNRS, Observatoire de la C\^ote d'Azur\\
Laboratoire J.L. Lagrange, Boulevard de l'Observatoire, CS  34229, 06304 Nice Cedex 4, France}

\author{T. Passot}
\affiliation{Universit\'e C\^ote d'Azur, CNRS, Observatoire de la C\^ote d'Azur\\
	Laboratoire J.L. Lagrange, Boulevard de l'Observatoire, CS  34229, 06304 Nice Cedex 4, France}

\author{P.L. Sulem}
\affiliation{Universit\'e C\^ote d'Azur, CNRS, Observatoire de la C\^ote d'Azur\\
	Laboratoire J.L. Lagrange, Boulevard de l'Observatoire, CS  34229, 06304 Nice Cedex 4, France}

\begin{abstract}
{A pair of nonlinear diffusion equations in Fourier space} is used to study the dynamics of strong Alfv\'en-wave turbulence, from MHD to electron scales. Special attention is paid to the regime of imbalance between the energies of counter-propagating waves commonly observed in the solar wind (SW), especially in regions relatively close to the Sun. In the collisionless regime where dispersive effects arise at  scales comparable to or larger than those where dissipation becomes effective, the  imbalance produced by  a given injection rate of generalized cross-helicity (GCH), which is an invariant, is much larger than in the corresponding collisional regime described by the usual (or reduced) magnetohydrodynamics.
The combined effect of high imbalance and ion Landau damping induces  a  steep  energy spectrum for the transverse magnetic field at sub-ion scales. This spectrum is consistent with observations in highly Alfvenic regions of the SW, such as trailing edges, but does not take the form of a transition range continued at smaller scales by a shallower spectrum.
This suggests that the observed spectra displaying such a  transition result from the superposition of contributions originating from  various streams with different degrees of imbalance. {Furthermore, when imbalanced energy injection is supplemented at small scales in an already fully developed turbulence, for example under the effect of magnetic reconnection, a significant enhancement of the  imbalance at all scales is observed.}
\end{abstract}



\section{Introduction}

An important contribution to the SW dynamics originates from nonlinear interactions between counter- propagating Alfv\'en waves. Outgoing waves are emitted at the Sun's surface as a consequence of mechanisms such as reconnection in the chromospheric magnetic network (see e.g. the recent review by \citet{Marsch18}), which leads to injection of Alfvenic waves  at the base of the fast SW \citep{McIntosh12}. Ingoing waves are created  by reflection on density gradients \citep{Chandran09,Chandran-Perez19} and by velocity shear \citep{Robert92,Breech08}, or result from parametric decay instability \citep{Vinas91,DelZanna01}. The energies carried by these counter-propagating waves are usually unequal \citep{Tu89,Lucek98,Wicks13}, with a degree of "imbalance" depending on the type of wind \citep{Tu90, Bruno14, Bruno17,DAmicis19} and also on the distance from the Sun \citep{Roberts87,Marsch-Tu90}.
Accurate in situ observations are now available for the turbulent energy cascades of Alfv\'en waves (AWs) or, at the sub-ion scales, kinetic Alfv\'en waves (KAWs) (see e.g. \citet{Kiyani15, Goldstein15} for reviews). A precise understanding of how imbalance affects these cascades is however required in order to predict how much turbulence can heat and accelerate the SW plasma \citep{Chandran10, Cranmer15, Mallet19}.
{Imbalance can also affect cosmic ray scattering efficiency, as for example discussed in \citet{Weidl15}}.

While in the MHD range, imbalanced Alfvenic turbulence has been extensively
studied both theoretically and numerically
(see \cite{Chen16} for review), its dynamics at sub-ion scales remains
largely unexplored. 
Due to the computational cost of 3D imbalanced kinetic simulations, one is led  to develop asymptotic models, isolating the AW dynamics  within a spectral range extending from the  MHD scales (larger than  ion and sonic Larmor radii) to the sub-ion scales, assuming a proton-electron homogeneous  plasma subject to a strong ambient magnetic field. Such a description is provided by a Hamiltonian two-field gyrofluid model retaining ion finite Larmor radius (FLR) corrections, parallel magnetic fluctuations and electron inertia  \citep{PST18}. It takes the form of two dynamical equations for the electron-gyrocenter density and the parallel magnetic potential, from which the electrostatic potential and the parallel magnetic fluctuations are easily derived. 
Numerical simulations of the two-field gyrofluid are in progress. Here we concentrate on a  {phenemenological} reduction of this model in the form of nonlinear diffusion equations in Fourier space for the transverse spectra of the two conserved quantities, energy and GCH \citep{PS19}. Concentrating on the effect of imbalance at scales larger than the electron skin depth $d_e$, electron inertia is neglected in most of the simulations, which all address the strong turbulence regime. 

The model involves
drastic simplifications, leaving for future study the influence of effects such as inhomogeneities and wind expansion.  In some of the simulations, we nevertheless retain Landau damping which was shown to affect the sub-ion inertial range of balanced AW  turbulence \citep{PS15, Sulem2016}, but neglect ion cyclotron damping and heating of the medium, considered for example by \cite{Cranmer03}, in spite of their potential effect on the dynamics.

\section{The model}
 
{In this section we provide a purely phenomonological derivation of the diffusion model discussed in \citet{PS19} for the time evolution of the transverse spectra  $E(k_\perp,t)$ and $E_C(k_\perp,t)$ of energy and GCH respectively.} The latter quantities are related to the energy spectra $E^\pm(k_\perp,t)$ of the counter-propagating waves by $E(k_\perp,t) = E^+ (k_\perp,t)+ E^- (k_\perp,t)$ and $E_C(k_\perp,t) = (E^+ (k_\perp,t)- E^- (k_\perp,t))/v_{ph}(k_\perp)$, where the (parallel) Alfv\'en phase velocity  $v_{ph}$ is a function of $k_\perp$.

{A nonlinear diffusion model suitable for three-wave interactions (dominant for KAW turbulence) and preserving  the existence of  absolute-equilibrium solutions should take the form \citep{Thalabard15}}
\begin{equation}\label{davidextended}
\frac{\partial }{\partial t} \frac{E^\pm(k_\perp)}{2}= \frac{\partial }{\partial k_\perp} \Big\lbrack D^\mp(E^+, E^-,k_\perp) \frac{\partial }{\partial k_\perp }\frac{E^\pm(k_\perp)}{k_\perp} \Big\rbrack + X^\pm ,
\end{equation}
where the corrective  terms $X^\pm$ are necessary because the energies of the forward and backward propagating waves are not  conserved independently in the presence of dispersion. Due to energy  conservation, it is clear that $X^+ = -X^-$. The equation for the GCH spectrum then reads
\begin{eqnarray}
\frac{\partial}{\partial t} \frac{E_C}{2} &=& \frac{2X^+}{v_\text{ph}} + \Big(\frac{\partial }{\partial k_\perp}+\frac{1}{v_\text{ph}}\frac{\partial v_\text{ph}}{\partial k_\perp}\Big) \times\nonumber \\
 \Big \{ &\sum_{r=\pm 1}& (-1)^{\frac{r-1}{2}}  \frac{D^{(-r)}}{v_\text{ph}} \frac{\partial}{\partial {k_\perp}} \Big(\frac{E^{(r)}}{k_\perp}\Big)\Big\}.\label{GCHconservation}
\end{eqnarray}
The invariance of the GCH, which implies that the rhs of~\eqref{GCHconservation} should be in a conservative form, allows us to find $X^+$ by the condition that only the first term in the round parentheses survives. Furthermore, dimensional analysis of the terms involved in Eq. \eqref{davidextended}, allows us to estimate $D^\mp \sim {k_\perp^3}/{\tau^\mp_\text{tr} }$ in terms of the transfer time $\tau^\mp_\text{tr}=(\tau^\pm_\text{NL})^2/\tau^\mp_\text{w}$. The case with Landau damping is treated in Appendix~\ref{Landau-damping}. In its absence, 
$D^\mp(E^\mp,k_\perp) =  C' k_\perp^6 \,v_\text{ph} E^\mp/{\widetilde k}_\|^{\mp}$,
where ${\widetilde k}_\|^{\mp}$ measures the inverse parallel correlation length of $\pm$ eddies or wave packets with perpendicular wavenumber $k_\perp$.
Here, we used arguments from \cite{PS19} to fix the nonlinear time $\tau^\pm_\text{NL}=(k^3_\perp v_\text{ph}^2 {E}^\pm)^{-1/2}$, consistently with the imbalanced strong MHD turbulence (see Lithwick et al 2007), and also wrote $\tau^\mp_\text{w}={{\widetilde k}_\|^{\mp}}v_\text{ph}$. {The parameter $C^\prime$ can be scaled out in the absence of Landau damping but in its presence affects the location of the transition to the exponential decay \citep{PS15}. In the following, it is chosen equal to unity.}  Retaining only strongly local interactions, the model ignores coupling between co-propagating waves since they have nearly the same velocity and can hardly overtake one another and interact. Interactions between such waves with comparable but not quasi-equal wavenumbers were considered by \citet{Voitenko16}, but they were shown not to significantly affect the dynamics in the context of the present model \citep{PS19}.

Assuming that both waves undergo a strong cascade $\tau^\pm_\text{NL}\sim \tau^\mp_\text{w}$, we get  ${\widetilde k}_\|^{\pm} = (k_\perp^3 {E}^\pm)^{1/2}$. However in MHD  it is  expected that in the imbalanced case, the $+$ wave (which by definition is more energetic) will undergo a weaker cascade as suggested by~\citet{Chandran08}. This effect can be modeled by changing the definition of the parallel wavenumber ${\widetilde k}_\|^-$  of the ``--'' wave  subject to interactions with the ``+'' wave. This suggests to introduce the parameter $\chi$ along with the following ansatz  
\begin{equation}
{\widetilde k}_\|^{(r)}=(k_\perp^3 {E}^{(r)})^{1/2}({E}^+/{E}^-)^{(1-r)\chi/4}. \label{kpar}
\end{equation}

This expression reproduces the different models found in the literature {for imbalanced MHD turbulence}, depending on the free exponent $\chi$. The value $\chi=0$ corresponds to the model of \cite{Lithwick07} \footnote{In \citet{Lithwick07}, a slightly different interpretation of the critical balance is used where instead of the wave period,  the correlation time of the straining
imposed by oppositely directed waves is considered. The associated correlation lengths differ from those of our model but the predicted spectra are the same.} where both  waves are in a strong turbulent regime and satisfy the critical balance condition. In contrast, $\chi=1$ reproduces the model of \citet{Chandran08}, where the transfer time of the stronger wave obeys a weak turbulence phenomenology. The value $\chi=1/4$ corresponds to the model of \citet{Beresnyak08}. {Numerical simulations of imbalanced MHD by these authors seem to favor a value of $\chi$ larger than zero but also strictly smaller than unity.
Differently, in the presence of dispersion, any value of $\chi$ smaller than one leads  to an unphysical instability at the dissipation scale \citep{PS19}.
The simulations without Landau damping described below (except the purely MHD ones) are thus performed with $\chi = 1$. The stability range is extended to $\chi \ge 0.5$ when Landau damping is retained.}

Finally, the set of reduced equations are cast as 
\begin{eqnarray}\label{masterequation}
&&\frac{\partial}{\partial t} \frac{E}{2} =  \frac{\partial}{\partial {k_\perp}}
\Bigg \{ k^{6}_\perp V\,\sum_{r=\pm 1} \frac{E^{(-r)}}{{\widetilde k}^{(-r)}_\|} \frac{\partial}{\partial {k_\perp}} \Big(\frac{E^{(r)}}{k_\perp}\Big)\Bigg\}
\label{strong-energy}\\
&&\frac{\partial}{\partial t} \frac{E_C}{2} =  \frac{\partial}{\partial {k_\perp}}
\Big \{ k^{6}_\perp\sum_{r=\pm 1} \frac{(-1)^{\frac{r-1}{2}} V E^{(-r)} }{v_{ph}{{\widetilde k}^{(-r)}_\|}} \frac{\partial}{\partial {k_\perp}} \Big(\frac{E^{(r)}}{k_\perp}\Big)\Big\}.
\label{strong-helicity}\nonumber\\
\end{eqnarray}

In the absence of Landau damping $V =v_{ph}$ (which is the only quantity in the model that incorporates the kinetic effects). It is constant at wavenumbers small compared with the smallest of the inverse ion Larmor radius $\rho_i^{-1} = (\sqrt{2\tau}\rho_s)^{-1}$ (where $\tau = T_{0i}/T_{0e}$ is the ion to electron temperature ratio at equilibrium)  
and  the inverse sonic Larmor radius $\rho_s^{-1} = (\sqrt{m_i/T_{oe}} \Omega_i)^{-1}$ (where $m_i$ is the ion mass and $\Omega_i$ the ion gyrofrequency), used as wavenumber unit
(with $\Omega_i^{-1}$ taken as time unit). It grows linearly 
($v_{ph} \approx \alpha k_\perp$) at smaller scales down to  $d_e=(2/\beta_e)^{1/2}\delta\rho_s$ (with $\delta^2$ denoting  the electron to proton mass ratio), where saturation occurs due to electron inertia. 

We treat Landau damping (see Appendix \ref{Landau-damping}) by adding dissipative terms  $- \gamma E$ and $-\gamma E_C$ in  Eqs. (\ref{strong-energy})-(\ref{strong-helicity}) respectively, where $\gamma(k_\perp, {\widetilde k}_\|)$ is the AW Landau dissipation rate  (dominated by electrons) of transverse and parallel wavenumbers
$k_\perp$ and ${\widetilde k}_\|= {\rm max}({\widetilde k}_\|^+, {\widetilde k}_\|^-)$. It also  affects (mostly through the ions) the transfer time \citep{PS15} and thus modifies the function $V$ entering Eqs. (\ref{strong-energy})-(\ref{strong-helicity}), which becomes $V= v_{ph}^2/(v_{ph}+ \mu \sqrt{2\tau})$ where $\mu$ is a numerical constant (in the following, $\mu=3$). 
\begin{figure}
	\includegraphics[width=0.5\textwidth]{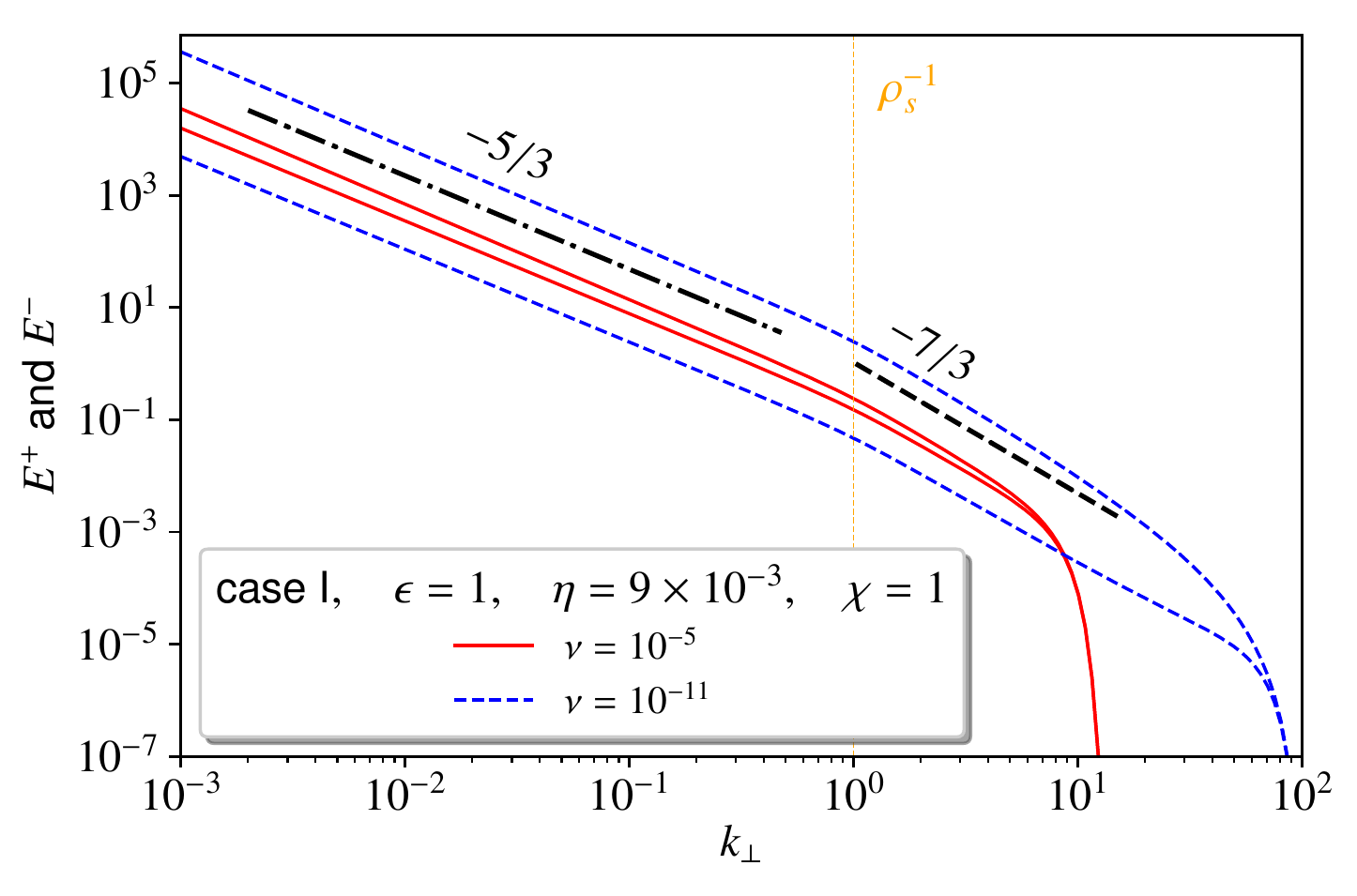}
	\includegraphics[width=0.5\textwidth]{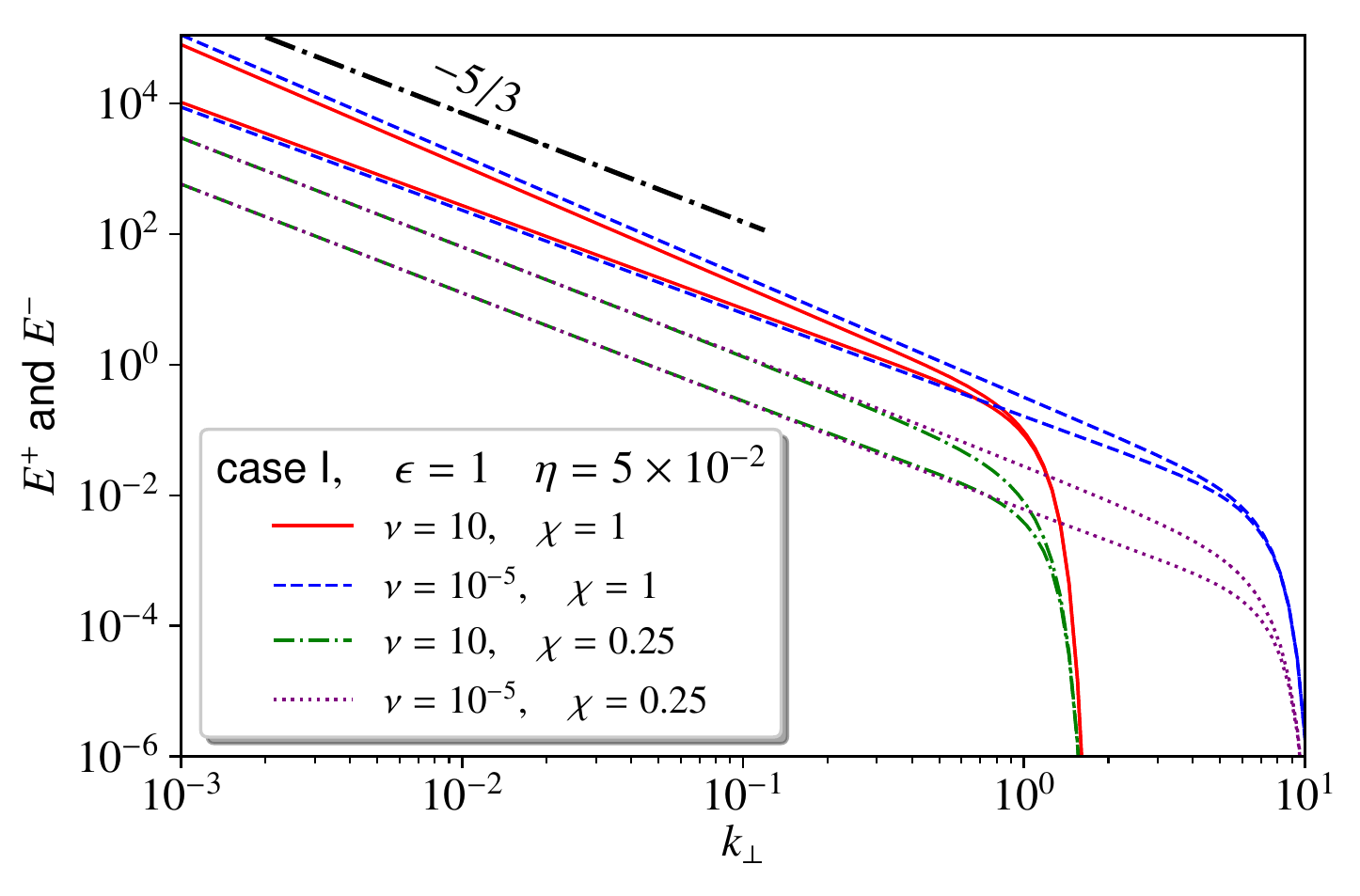}
	\caption{Changes in the imbalance between the $E^\pm$ spectra when, for prescribed energy and GCH transfer rates, the dissipation scale is varied via different values of $\nu$.  Top: fully dispersive equations. Bottom: pure MHD regime, where  in the $\chi = 0.25$ run, $E^\pm$ are divided by factor 10 for better clarity. Here  and in  similar further figures, the same color and line style are used for spectra $E^+$ and $E^-$ corresponding to the same run.}\label{imbalanceVSk_d}
\end{figure}

\begin{figure}
\includegraphics[width=0.5\textwidth]{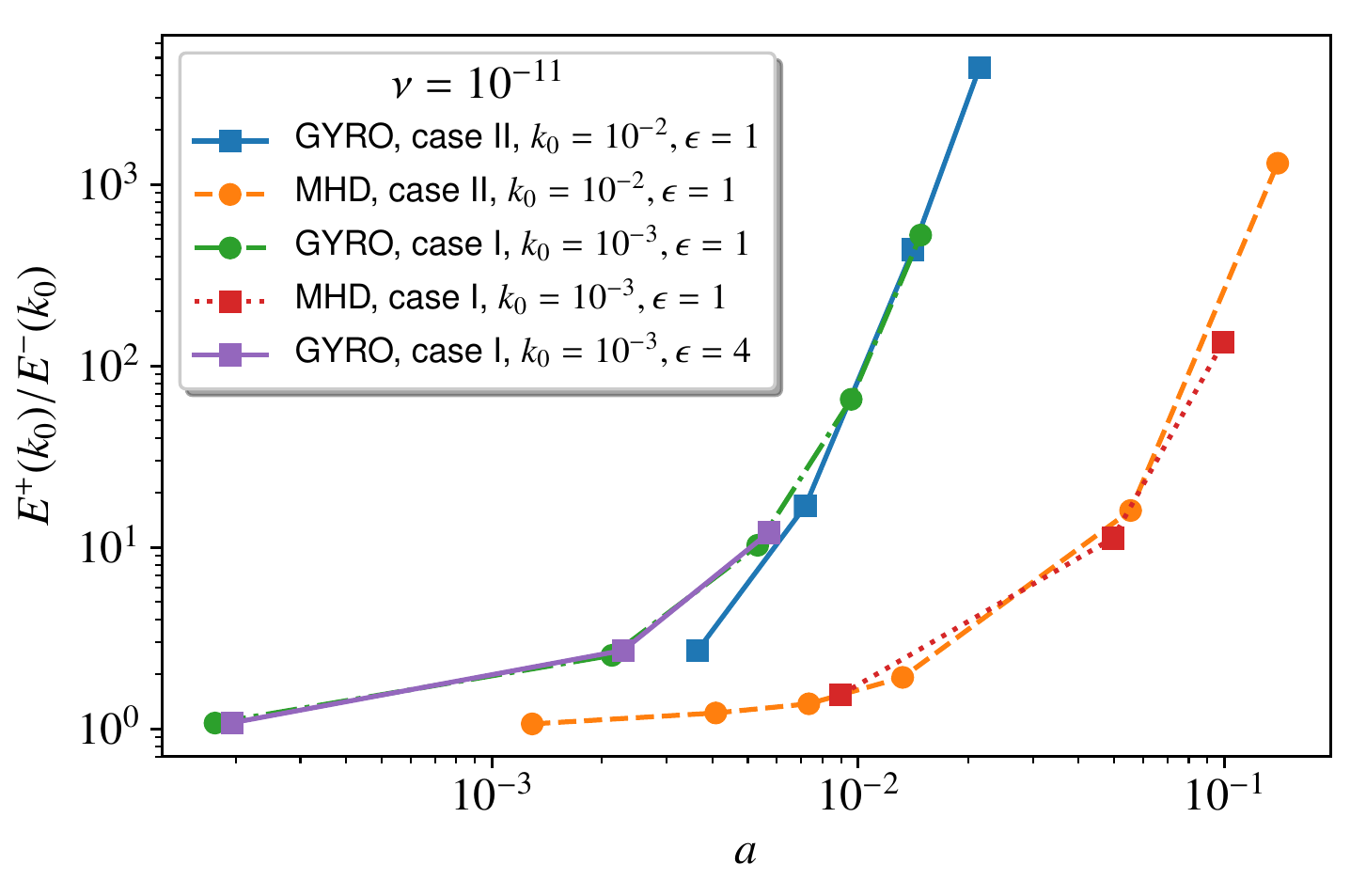}
\caption{Imbalance at the outer scale $k_0$ versus  $a=\sqrt{2/\beta_e}\eta/\epsilon$ (MHD)  and $a=\alpha \eta/\epsilon$  (GYRO), with $\chi=1$.\label{SetofimbVScross}
}
\end{figure}

Equations (\ref{strong-energy})-(\ref{strong-helicity}) were solved numerically using a finite difference scheme both in $k_\perp$ (with a logarithmic discretization) and $t$ (using  a forward Euler scheme with  adaptive time step),  modifying a code developed for  weak turbulence of gravitational waves \citep{GALTIER201984}.   We also  introduced different grids for the fields and the fluxes, with linear interpolation on the fields. Their coupling ensures better stability. 

Simulations were performed for $\beta_e = 2,\tau = 1$ (case I) typical of the SW at  1 AU, and $\beta_e=0.04, \tau=10$ (case II), more suitable for regions closer to  the Sun \citep{Roytershteyn_2019}. For small  $\beta_e$, $d_e$ significantly exceeds the electron Larmor radius $\rho_e = \sqrt{2} \delta \rho_s$, thus permitting  electron inertia to be retained, while electron FLR corrections are neglected. Simulations
with and without Landau damping were performed, the latter regime being of interest for comparison with analytical predictions. In this case,  hyperdiffusive terms of the form $\nu k_\perp^8 E(k_\perp,t)$
and $\nu k_\perp^8 E_C(k_\perp,t)$, are supplemented in Eqs. (\ref{strong-energy})-(\ref{strong-helicity}),
with a coefficient $\nu$ depending on the resolution and the parameters. The system is driven at large scale (injection wavenumber $k_0$), dissipation taking place at small scales (dissipation or pinning wavenumber $k_d$ between $10^{1}$  and $10^{4}$). Driving is performed either through boundary conditions by prescribing the  spectra $E^\pm$  or the fluxes $\eta$ and $\epsilon$ of energy and GCH at the smallest wavenumber $k_{\rm min} = k_0$ (within the range $10^{-3}$ to $10^{-2}$), or differently through injection terms  of the form $\varepsilon f(k_\perp)$ and $\eta f(k_\perp)$ supplemented in the r.h.s. of Eqs. (\ref{strong-energy})-(\ref{strong-helicity}), $f$ denoting a function with a compact support located near the injection wavenumber such that $\int_0^\infty f(k_\perp) dk_\perp=1$. Initial conditions consist of a spectral bump in the MHD range near the smallest retained wavenumber.

\section{{Direct cascades in imbalanced turbulence}}\label{analytics}

\begin{figure}
	\includegraphics[width=0.5\textwidth]{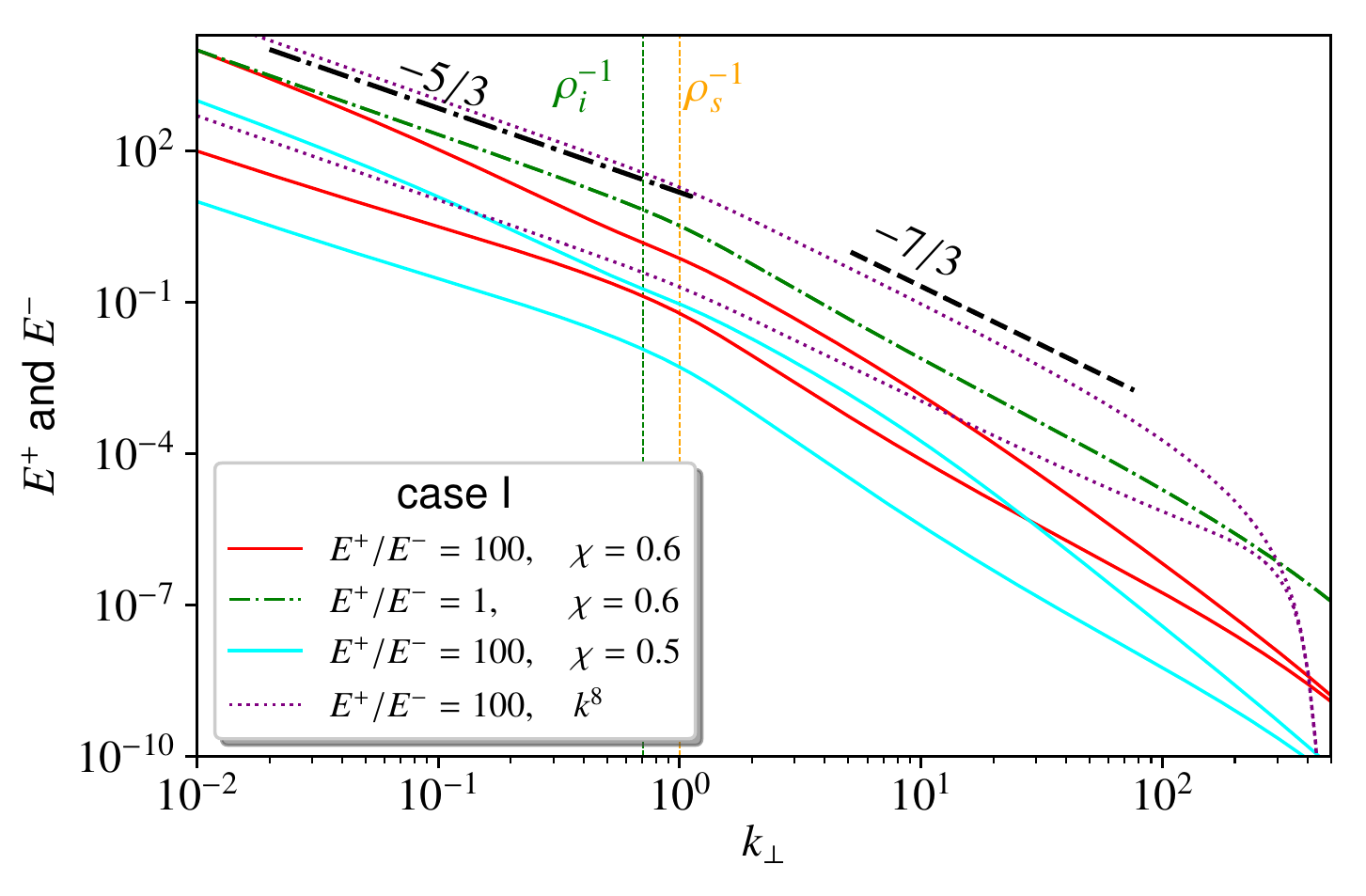}
	\includegraphics[width=0.5\textwidth]{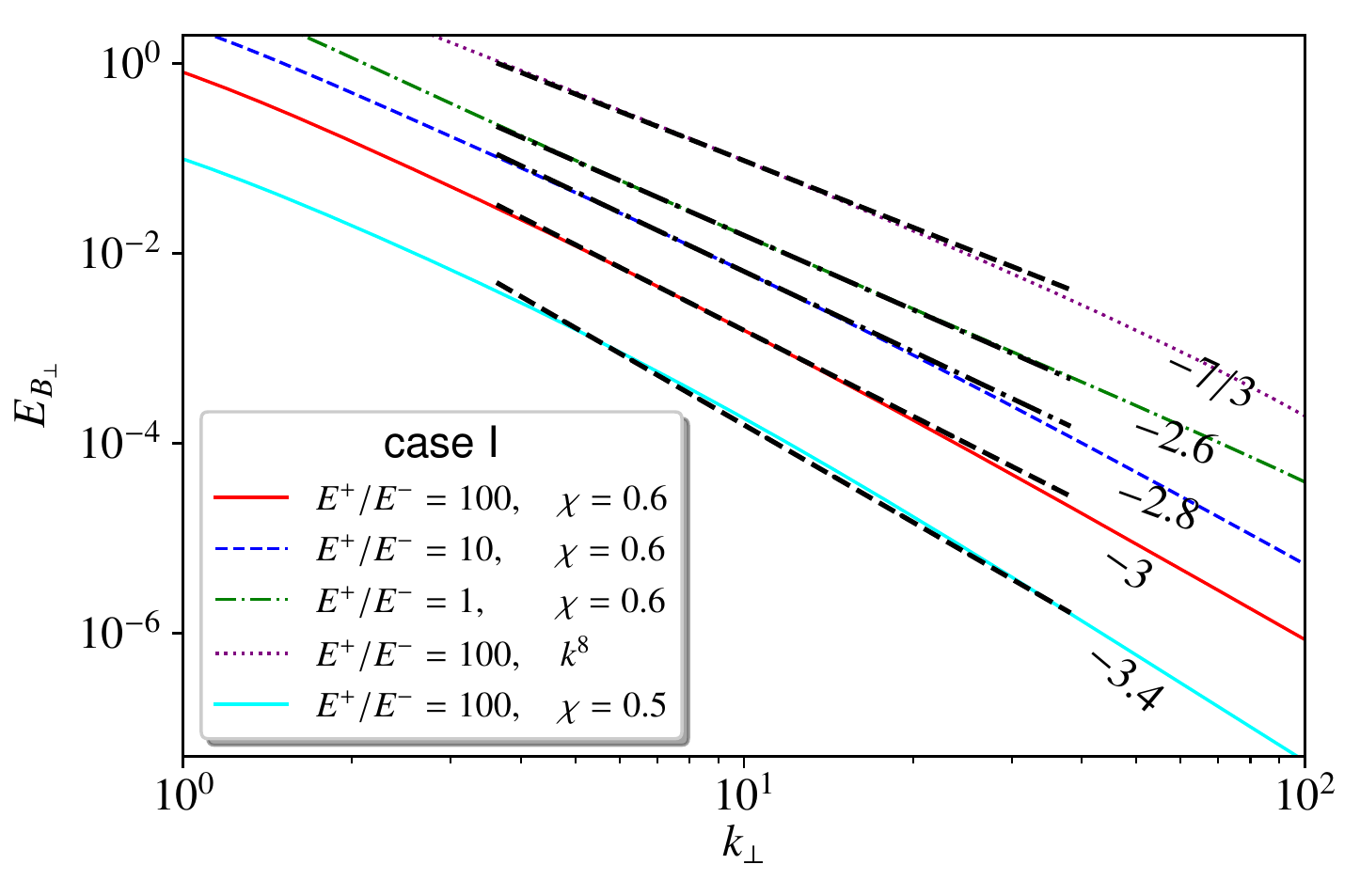}
	\caption{Top: spectra $E^\pm$  when prescribed at the outer wavenumber $k_0=10^{-2}$,  for four simulations ($\delta=0$): Landau damping is used in all simulations except when labeled by $k^8$ . Bottom: Sub-ion $E_{B_\perp}$ spectrum. Curves are shifted vertically for better readability.} \label{landaudamping}
\end{figure}

\begin{figure}
	\includegraphics[width=0.5\textwidth]{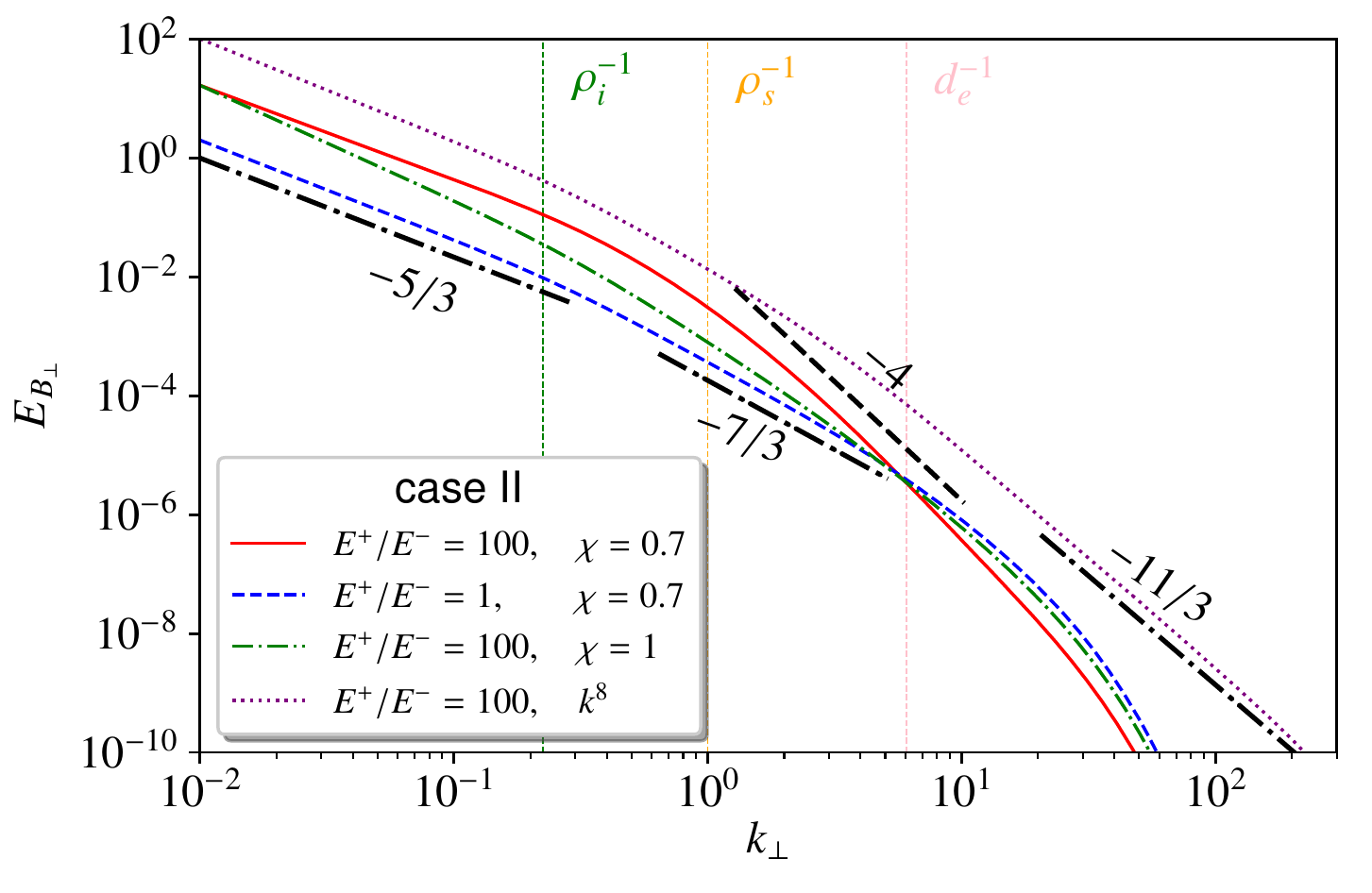}
\caption{$E_{B_\perp}$ spectrum ($\delta^2=1/1836$). As  imbalance is increased,  spectra get steeper  in the presence of Landau damping than 
	with $k^8$-hyperdiffusion.}
\label{landausteepnening}
\end{figure}
When, in the absence of  Landau damping,  Eqs. (\ref{strong-energy})-(\ref{strong-helicity}) are integrated in the strong turbulence regime with prescribed boundary values $E^\pm(k_0)$ and $k_\perp^8$ dissipation terms, we observe a direct transfer of energy and GCH, with the development of power-law spectra which progressively extend to the small scales, up to the moment when the dissipation scale is reached. Although the MHD-range spectrum that forms during this transient is very close to the $k_\perp^{-5/3}$ stationary solution, the sub-ion spectra  first develop a balanced regime (as a result of a faster transfer of energy than of GCH) with a spectrum steeper than $k_\perp^{-8/3}$ before converging, while imbalance develops, towards an approximately $-7/3$ steady state when imbalance is weak (not shown).  Nonlinear diffusion models for finite-capacity systems (no ultra-violet divergence of the energy) often exhibit  such transient anomalous  power-law spectra \citep{Thalabard15}. The case of weak KAW turbulence has recently been addressed by \citet{Galtier19}.

When the boundary conditions in  $k_0$ consist of prescribing energy $\epsilon$ and GCH $\eta$  fluxes, the imbalance between  $E^+$ and $E^-$ at large scales is found to strongly depend on the small-scale extension of the dispersive range, {as seen in Fig.~\ref{imbalanceVSk_d} (top). This is to be compared with  Fig.~\ref{imbalanceVSk_d} (bottom) where the phase velocity is assumed constant (non-dispersive MHD). Both $\chi=1$ (for comparison with the dispersive case), and the more realistic value $\chi=0.25$ are displayed.}  {In both panels, the spectra are plotted at times at which a stationary regime with constant (positive) energy and GCH fluxes have established.} In these simulations (case I with $\delta=0$), we
varied the dissipation wavenumber by changing the hyperdiffusivity coefficient. In the MHD regime,  $E^\pm$ behave as power laws until the pinning scale is approached. Conversely, in the dispersive case, the spectra approach each other exponentially (consistent with the linear variation of  $\phi(k_\perp)=(1/2)\ln (E^+(k_\perp)/E^-(k_\perp))$ when $\alpha\eta/\epsilon \ll 1$ (see Appendix \ref{imbalance}). It follows that, in contrast with usual hydrodynamic turbulence, large-scale quantities depend on the dissipation scale $k_d^{-1}$. This effect is much stronger in the presence of dispersion where $E^+(k_0)/E^-(k_0)$ varies exponentially with $k_d$, while it scales like a power law in  MHD {when keeping $\chi=1$} (see Eqs. (\ref{imbalancegyro}) and (\ref{imbalancemhd}) respectively). {For MHD with $\chi=0.25$, the spectra differ by less than $1\%$ from the case $\chi=0$ for which there is no dependence on $k_d$ (see Eq. (\ref{imbalance-chi=0})).}
{To understand heuristically the behavior in the dispersive case, we must note that (i) the dissipative scales for $E$ and $E_C$ are the same since the equation for the GCH spectrum is linear in $E_C$,  (ii) the $E_C$ spectrum steepens at the dispersive scale even more than the energy spectrum since $|E_C(k_\perp)|\le E(k_\perp)/v_{ph}(k_\perp)$. As a consequence, in order for 
	the GCH dissipation $\nu \int k_\perp^8 E_C dk_\perp$ to match the injected GCH rate $\eta$ prescribed at $k_0$, the magnitude of $E_C$  must be larger, and consequently the imbalance at large scales enhanced compared with the MHD problem in the same setting. }

Difference between the dispersive (GYRO) and non-dispersive (MHD) cases is also seen on Fig.~\ref{SetofimbVScross} which displays the  imbalance $E^+(k_0)/E^-(k_0)$ at the outer wavenumber $k_0=k_{\rm min}$ (located in the MHD range) versus the parameter $a\equiv v_{ph}(k_0)\eta/\epsilon\approx \sqrt{2/\beta_e}\eta/\epsilon$ (MHD) or $a\equiv\alpha\eta/\epsilon$ (GYRO), with $\alpha$ defined in Appendix \ref{imbalance}, in cases I and II.  Graphs corresponding to different $\beta_e$ collapse on the same curve, with an almost perfect agreement in MHD. Note that, in the GYRO simulations, the assumption $v_{ph}=\alpha k_\perp$ is only approximate, especially close to the ion scale.
We furthermore observed that changing $\varepsilon$ while keeping  $\eta/\varepsilon$ constant has no effect. 

Influence of  the degree of imbalance in the presence of Landau damping is considered in Fig.~\ref{landaudamping} in  case I with $\delta=0$, where, for  comparison, a simulation involving hyperdiffusion is also presented.  Top panel displays the energy spectra $E^\pm$, while  bottom panel shows the transverse magnetic spectrum $E_{B_\perp}$ (see Appendix \ref{Bspectrum}). With hyperdiffusion, the  $k_\perp^{-5/3}$  MHD spectrum is continued at sub-ion scales by the classical $k_\perp^{-7/3}$ range, the degree of imbalance significantly decreasing only near the pinning wavenumber. When Landau damping is retained,  the degree of imbalance decreases with the scale, the more so when $\chi$ is closer to $1$ (visible in the top panel when comparing the runs with $E^+(k_0)/E^-(k_0) = 100$ for $\chi=0.6$ and $0.5$).  In the sub-ion range, the steepening of $E_{B_\perp}$ displayed in the bottom panel  increases with the degree of imbalance (a more pronounced effect when $\chi$ is smaller). Such a steepening is often observed  as a transition range in the SW at 1AU, depending on the fluctuation power \citep{Bruno14}, and sometimes associated with proton Landau damping \citep{Sahraoui10} and imbalance degree or Alfvenicity \citep{Bruno14,Bruno17,DAmicis19}.  The present model suggests that both effects are to act simultaneously.  An alternative mechanism for steepening of the spectrum related to reconnection is suggested  by \citet{Vech18}.

Figure~\ref{landausteepnening} displays $E_{B_\perp}$ in case II with $\delta^2 = 1/1836$. For this relatively small $\beta_e$, an even stronger  spectral steepening is observed for large imbalance. Furthermore, the multiplicative factor in the estimate of $E(k_\perp)$ in terms of $E_{B_\perp}(k_\perp)$ is responsible for a steepening of the latter at scales smaller than $d_e$, an effect visible in particular in the simulation with hyper-diffusivity, $\chi=1$,  and $E^+(k_0)/E^-(k_0)=100$. A $k^{-11/3}_\perp$ spectrum, classical for balanced IKAW turbulence \citep{ChenBold17,PST18,Roytershteyn_2019} is still observed for this level of imbalance. Spectra, obtained with Landau damping and electron inertia, are also  displayed in this figure: for $\chi=0.6$ with $E^+(k_0)/E^-(k_0)=1$ or $E^+(k_0)/E^-(k_0)=100$, showing that the spectra get steeper as the imbalance increases, and for $\chi=1$ with $E^+(k_0)/E^-(k_0)=100$, showing that the steepening is more pronounced as $\chi$ is decreased. Such a steepening (although more moderate) is also observed in 3D fully-kinetic  simulations~\citep{Groselj18}.

\section{Inverse transfer of imbalance}

\begin{figure}
\includegraphics[width=0.5\textwidth]{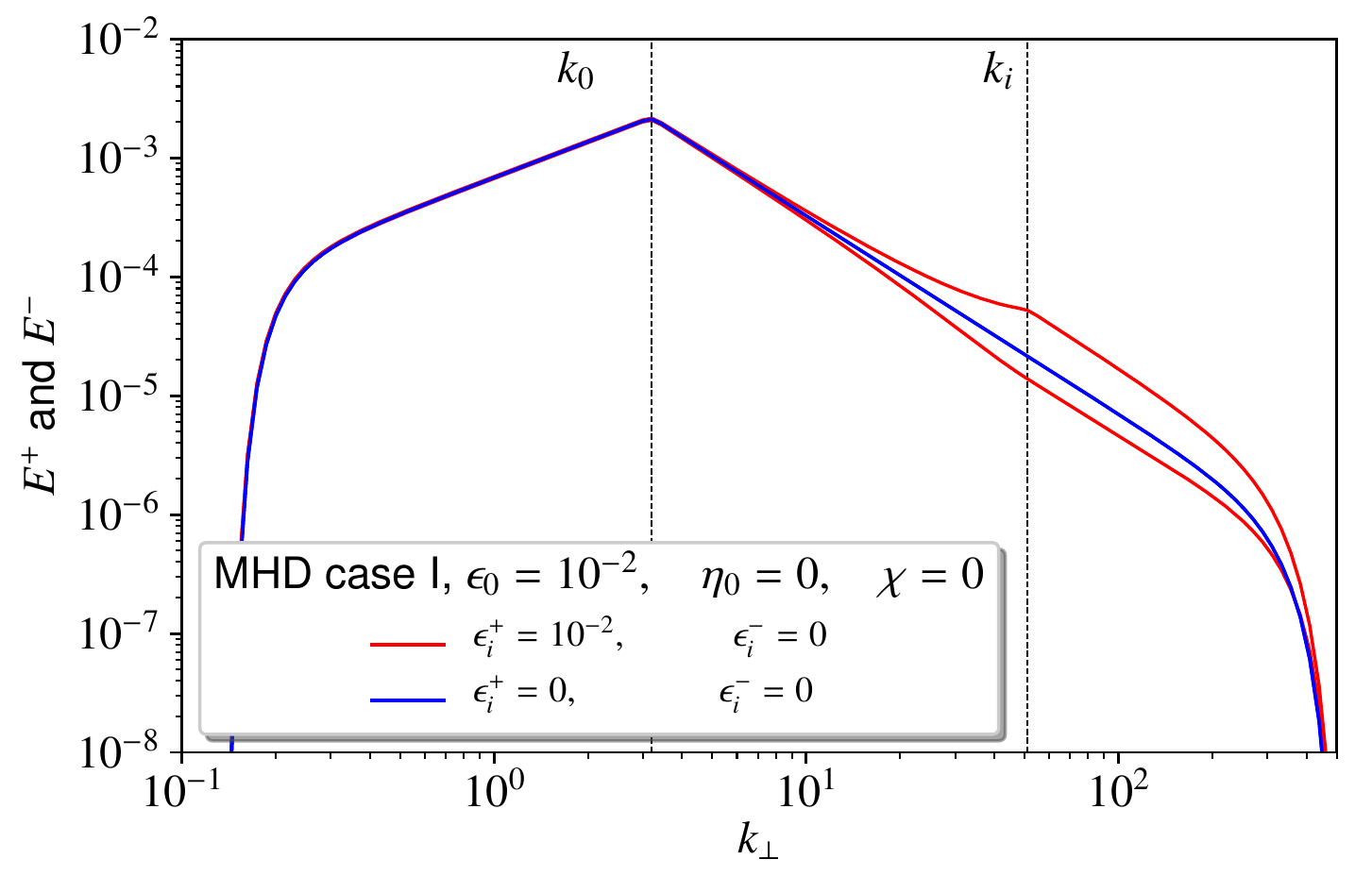}
\includegraphics[width=0.5\textwidth]{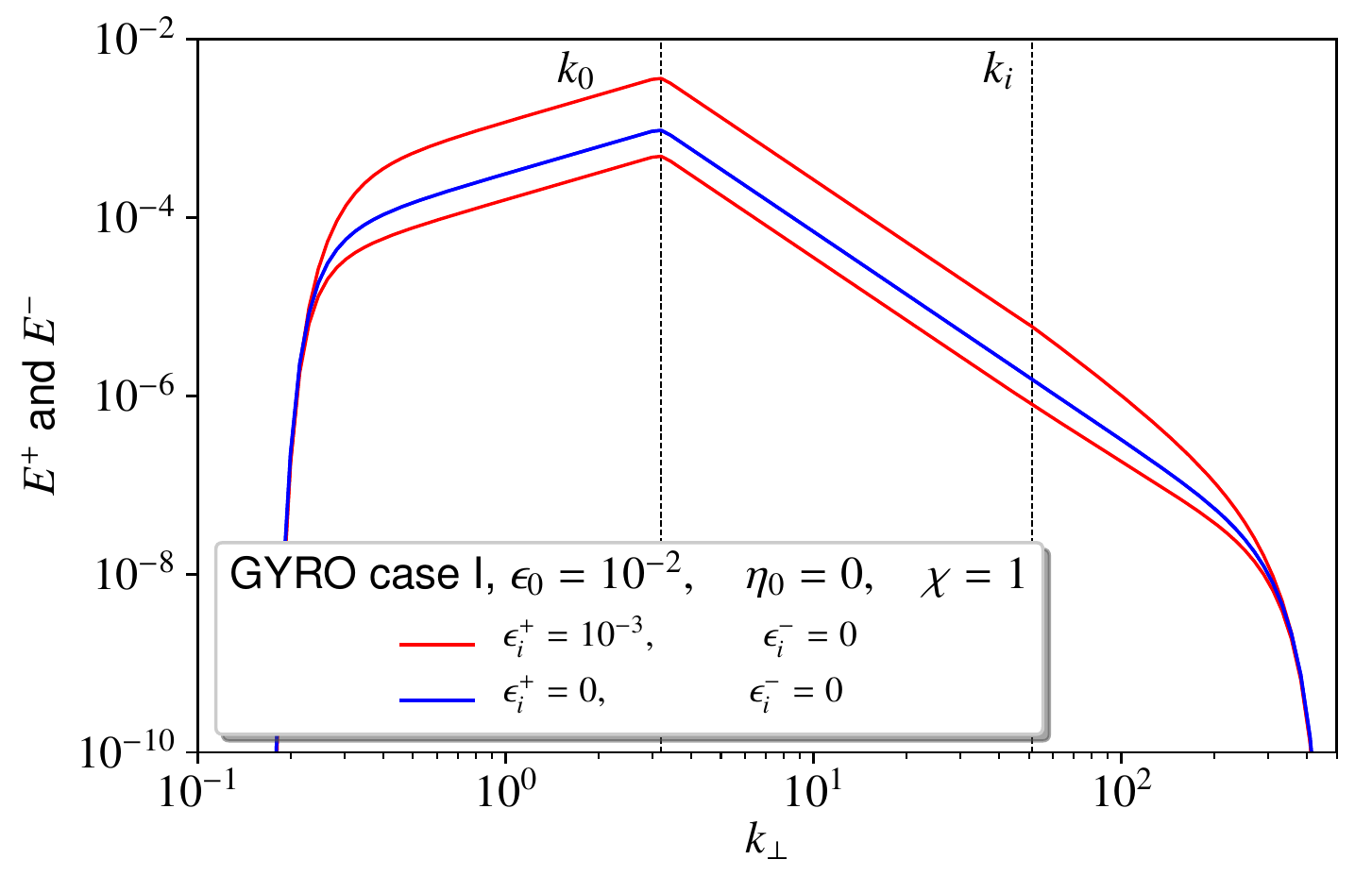}
\caption{$E^\pm$  spectra in MHD (top) and GYRO (bottom)  simulations when GCH is injected at  $k_i \gg k_0$ with a rate $\eta_i$, in a developed turbulence without Landau damping driven  at  $k_0$ with rates $\epsilon_0$ and $\eta_0$. 
}
\label{inversecascade}
\end{figure}

{We consider a situation where a stationary turbulence is affected by small-scale injection of KAWs resulting e.g. from magnetic reconnection (\citet{Chaston05,Liang16,Shi19} and references therein). We focus on  the case where the injection is imbalanced. This setting differs from that used for studying  a possible inverse cascade of GCH by driving an initially zero solution. Such a cascade, predicted to exist at sub-ion scales using absolute equilibrium arguments \citep{PST18}, and necessarily involving inverse energy transfer, is not captured by the diffusion model and is currently investigated via direct simulations of the gyrofluid model.}

{In the following, the initial stationary regime is obtained with a large-scale driving at $k_0=3.2$ (larger than $k_\text{min}$) and a $k^8$ hyperdiffusivity  $\nu=10^{-16}$ in GYRO and $\nu=10^{-18}$ in MHD runs. An injection of $E^+$ energy only (maximal imbalance) is then provided at a wavenumber $k_i=51.2$ at a rate smaller than or equal to the one imposed at large scales. Stationary solutions are obtained by using a $k^{-8}$ hypodiffusivity equal to $10^{-7}$ for MHD and $10^{-6}$ for GYRO.   The results presented below pertain to the case where the initial turbulence is balanced but later becomes imbalanced as a result of additional forcing at $k_i$. }

{Let us first address the regime where the driving takes place in the MHD range for $\chi=0$.
We see on Fig. \ref{inversecascade} (top) that some imbalance is generated at wavenumbers larger than $k_i$  but not at smaller wavenumbers, indicating the absence of inverse transfer of GCH.
This point can be understood by solving Eq. (\ref{eqphi-chi=0}) (right) for $\phi=(1/2)\log(E^+(k_\perp)/E^-(k_\perp))$ 
with initial condition taken at $k_0\ll k_\perp<k_i$. For the parameters of the simulation, the initial condition for $\phi$ is of order unity and the constant $C$ turns out to be very large, leading to a fast decrease of $\phi-\psi$ (where $\psi= 2\,{\tanh}^{-1} a$) with $k_\perp$. We thus find that for wavenumbers approaching $k_0$, the imbalance remains unchanged, equal to the value given by Eq. (\ref{imbalance-chi=0}). Numerical simulations show that the absence of inverse transfer of GCH still holds for larger values of $\chi$, including in particular $0.25$. }

{When injection takes place in the dispersive range (with $\chi=1$), assuming that $\alpha\eta/\epsilon\ll 1$, 	Eq. (\ref{eq:phi-dispesive}) indicates that $\phi$ is almost independent of $k_\perp$, and thus that the imbalance created at $k_i$ extends to large scales, as exemplified in Fig. \ref{inversecascade} (bottom).
This evolution results from the development of a transient negative GCH flux at $k_\perp<k_i$ due to the small-scale forcing, leading to a decrease of the total GCH flux and to an enhancement of imbalance that saturates when the flux recovers its original value (consistent with the GCH flux $\eta_0$ injected at $k_0$). In the SW, dispersive scales are always present, making the possibility of such an inverse transfer  relevant. It would be of interest to investigate whether the process by which reconnection events  can generate an inverse flux
towards larger scales \citep{Franci17} can also generate imbalance in the case where  KAWs are  generated at the reconnection sites. }
The present modeling could also be useful to analyze the recently predicted cascade reversal  at $d_e$ in a reduced two-fluid model~\citep{Miloshevich18} and a 3D extended magnetohydrodynamic model~\citep{Miloshevich17}.

\section{Conclusion}

This letter provides an analysis of the influence of imbalance between the energies of counter-propagating AWs, on the dynamics of a collisionless plasma. It is aimed  to contribute to the understanding of regimes encountered in the SW, particularly in regions close to the Sun explored by space missions such as Parker Solar Probe or Solar Orbiter.  

The main results can be summarized as follows. (i)  The imbalance produced by large-scale injection of GCH at a prescribed rate is enhanced by  wave dispersion. (ii) A steep range in the spectrum of the transverse magnetic fluctuations, consistent with the transition region  reported in SW observations \citep{Sahraoui10}, develops at the sub-ion scales, under the  combined influence of Landau damping and strong imbalance, an effect enhanced when $\beta_e$ is decreased. We conjecture that the  simulation results  can be more specifically related to the observations inside the trailing edge, which is characterized by the highest level of Alfvenicity, i.e. imbalance \citep{Bruno14}.  Existence of a shallower spectrum at smaller scales is then expected to originate from SW regions that are less imbalanced and  more energetic at these scales. (iii) Under some conditions the system develops an {inverse transfer of imbalance when imbalanced forcing takes place at small scales in an already fully developed turbulence.}

Future works include the study of the parent two-field gyrofluid models which is in particular expected to address the
question of the characteristic  nonlinear time scale in imbalanced turbulence, evaluate
the assumption of strongly local interactions and investigate the role of KAW decay instability. The influence on the global dynamics of the coupling of the AWs with the slow modes, important at small $\beta_e$ as they can generate large-scale parametric decay instabilities, will be studied using an extension of the present gyrofluid including both kinds of waves.

\acknowledgments{Computations have been done on the "Mesocentre SIGAMM" machine, hosted by Observatoire de la C\^ote d'Azur.}

\appendix
\section{Modeling Landau damping} \label{Landau-damping}
The dissipation rate $\gamma(k_\perp, k_\|)$ is evaluated from Eq. (D.21) of \citet{HCD06}, obtained from the  linearized gyrokinetic equations in the limit $\delta^2\tau\ll \beta_i=\tau\beta_e\ll 1$  (c.f. definitions in the text). In a non-dimensional form, one has 
\begin{eqnarray}\label{landaudissipation}
\gamma = \sqrt{\frac{\pi}{2}}\frac{1}{ \beta_e} \Big(\frac{\Gamma_0(\tau k_\perp^2)}{\tau^{3/2}} \exp{\Big\lbrack - \frac{\overline{\omega}^2}{\tau \beta_e}\Big\rbrack} + \delta\Big) k_\parallel k^2_\perp.
\end{eqnarray}
Here $\overline\omega^2 = k^2_\perp(1+\tau-\Gamma_0(\tau k^2_\perp))/({1-\Gamma_0(\tau k^2_\perp)})$, where $\Gamma_n(x) = I_n(x)e^{-x}$ and $I_ n$ is the first type modified Bessel function of order $n$. While Eq. ~\eqref{landaudissipation} includes both ion and electron Landau damping, at small $\beta_e$, the primary contribution comes from electrons, so that Eq. (63) of \citet{HCD06} can also be used.
Landau damping also affects the transfer times $\tau_{tr}^\pm$ of both counter-propagating waves, due to the temperature homogenization process along the magnetic field lines on the correlation length scale $k_\|^{\pm -1}$. The associate time scale $v_{th} {\widetilde k_\|^\pm}$,  which explicitly arises in Landau fluid closures \citep{Snyder97, SP15},    being proportional to the thermal velocity $v_{th}$ of the particles, is very short for the electrons and cannot affect the dynamics. It is in contrast relevant in the case of the ions for which it is given in the present units by $(\tau_{H}^\pm)^{-1}= \mu \sqrt{2 \tau}\, {\widetilde k_\|^\pm}$, where $\mu$ denotes a numerical constant of order unity.  
We are thus led to write  
\begin{equation}
\tau_{tr}^\pm = \tau_{NL}^\pm \left (\frac{\tau_{NL}^\pm}{\tau_w^\pm} + \frac{\tau_{NL}^\pm}{\tau_H^\pm} \right),
\end{equation}
which leads to
\begin{equation}
V = \frac{v_{ph}^2}{v_{ph} + \mu\sqrt{2\tau}}.
\end{equation}

\section{Imbalanced regime} \label{imbalance}
\setcounter{equation}{0}
\renewcommand{\theequation}{\ref{imbalance}.\arabic{equation}}
It is possible to relate the flux ratio $\eta/\epsilon$ to the imbalance $E^+(k_0)/E^-(k_0)$ at the outer scale. This can be done by rewriting the spectra in the form
\begin{eqnarray}
&&\frac{E(k_\perp)}{k_\perp}=\rho(k_\perp)\cosh\phi(k_\perp),\qquad\frac{v_{ph}(k_\perp)E_C(k_\perp)}{k_\perp}=\rho(k_\perp)\sinh\phi(k_\perp).
\end{eqnarray}
 This  leads to
\begin{equation}
E^\pm(k_\perp)=\frac{1}{2}k_\perp \rho(k_\perp)e^{\pm\phi(k_\perp)}.\label{Epmdefinition}
\end{equation}
Two cases are to be distinguished, depending on the value of $\chi$.
\subsection{The case $\chi=1$}
From Eq.~\eqref{masterequation} we see that, in the case $\chi = 1$, solutions with constant fluxes $\eta$ and $\epsilon$ obey 
\begin{eqnarray}
&& \frac{d}{d k_\perp}\rho^2(k_\perp)=-\frac{2\varepsilon \,{\widetilde k_\|^+(k_\perp)}}{k^7_\perp v_{ph}(k_\perp)}, \qquad \rho^2(k_\perp)\frac{d}{d k_\perp}\phi(k_\perp)=-\frac{\eta \,{\widetilde k_\|^+(k_\perp)}}{ k^7_\perp}, \label{rho-phi}
\end{eqnarray}
where ${\widetilde k_\|^+} = k_\perp^2 \sqrt{\rho/2}\exp (\phi/2)$. 

\begin{itemize}
	\item 	In the MHD regime where $v_{ph} = \sqrt{2/\beta_e}$, it is easily shown that, defining $a=\eta v_{ph}/\epsilon$, 
	\begin{equation}\label{varphimhd}
	\phi = \phi_0 + a\ln{\rho/\rho_0}.
	\end{equation} 
	When substituted into Eq. (\ref{rho-phi}, left),  this leads  to 
	$\rho = \Big((3-a)\,e^{\phi_0/2}\epsilon\, \sqrt{\beta_e}/16\Big)^{2/3} k_0^{-8/3}(k_\perp/k_0)^{-8/(3-a)},
	$
	which, after some  algebra,  prescribes for the MHD regime
	\begin{equation}
	\frac{E^+(k_0)}{E^-(k_0)} = \Big(\frac{k_d}{k_0}\Big)^{\frac{16a}{3-a}} , \quad \phi = \phi_0 - \frac{8 a}{3-a} \ln \frac{ k_\perp}{k_0},\label{imbalancemhd}
	\end{equation}
	where $k_d$ denotes the pinning wavenumber. This prediction excellently matches the numerical results presented in Fig.~\ref{SetofimbVScross}.
	\item In the far sub-ion range, $v_{ph}\approx \alpha k_\perp$, where 
	$\displaystyle{\alpha = 2  \sqrt{\frac{1+\tau}{\beta_e(2+(1+\tau)\beta_e)}}}$.
	The relationship between $\rho$ and $\phi$ derived from Eq. (\ref{rho-phi}) reads in this case
	\begin{equation}
	\frac{\alpha \eta}{\epsilon} k_\perp \frac{d}{dk_\perp} \ln{\rho} = \frac{d\phi}{d k_\perp}.\label{eqforphi}
	\end{equation}
	If we assume  $\eta \alpha / \epsilon \ll 1$,  $\phi$ can be approximated by a constant $\phi_0$  in Eq. (\ref{rho-phi}, left) which is then solved as
	\begin{equation}
	\rho = \Big(\frac{3\,\epsilon\, e^{\phi_0/2}\sqrt{\beta}}{10\alpha\sqrt{2}} \Big)^{2/3} k_\perp^{-10/3}.
	\end{equation}
	Equation ~\eqref{eqforphi} becomes
	\begin{equation}
	\phi = \phi'_0 - \frac{10}{3}\frac{\alpha\eta}{\epsilon}k_\perp, \label{eq:phi-dispesive}
	\end{equation}
	which determines the  pinning wavenumber $k_d$ where  $\phi$  vanishes. From Eq. ~\eqref{Epmdefinition}, it is clear that for small $\alpha \eta/\epsilon$,  $\phi$ will be nearly constant up to  the vicinity of  $k_d$  where the spectra $E^\pm$ approach  each other exponentially. Therefore, we can obtain a dispersive imbalance relation, whose behavior is very different from the MHD case,  namely
	\begin{equation}
	\frac{E^+(k_0)}{E^-(k_0)} = \exp \left (\frac{20 \alpha\eta (k_d-k_0)}{3\epsilon} \right ), \quad \phi'_0 =  \frac{10 \alpha\eta k_d}{3\epsilon} \quad\text{when}\quad k_d \gg k_0. \label{imbalancegyro}
	\end{equation}
	{To apply this formula to the case where both MHD and dispersive ranges are present, one has to match $\phi$ at the transition wavenumber $k_\perp$ using both~\eqref{imbalancemhd} and~\eqref{eq:phi-dispesive}. But the contribution due to the second term in~\eqref{imbalancemhd} is negligible and therefore we can simply extend~\eqref{imbalancegyro} to the full range.}
\end{itemize}

\subsection{The MHD regime with $\chi=0$} \label{MHD-chi0}
After some algebra, it is easy  shown that the equations for $\rho$ and $\phi$ read
\begin{equation}
\rho\frac{d}{d{k_\perp}} \left (\rho^{1/2}(k_\perp)\cosh(\frac{\phi(k_\perp)}{2})\right)=-\frac{\epsilon}{2^{3/2}k^5_\perp v_{ph}},\qquad \rho\frac{d}{d{k_\perp}} \left (\rho^{1/2}(k_\perp)\sinh(\frac{\phi(k_\perp)}{2})\right)=-\frac{\eta}{2^{3/2}k^5_\perp}.
\end{equation}
From here, defining $\tanh{\psi/2} = \eta v_{ph}/\epsilon$, we derive ($A$ being a constant),
\begin{equation}
\sqrt{\rho} \sinh{\frac{\phi-\psi}{2}} = A,\quad \frac{d\phi}{dk} = \frac{C}{k_\perp^5}\sinh^4 \frac{\phi - \psi}{2},
\label{eqphi-chi=0}
\end{equation}
where $\displaystyle{C=\frac{\epsilon/v_{ph}}{2^{1/2} \,  A^3 \cosh{\psi/2}}}$.
{Let us consider the case where energy and GCH are injected at $k_\perp=k_i$. Imposing $\rho=0$ as boundary condition at $k_\perp=+\infty$, we find that $A=0$. As a result $\phi = \psi$ for  $k_\perp > k_i$}. In this case, the imbalance becomes independent of $k_\perp$ in the form

\begin{equation}
\frac{E^+(k_\perp)}{E^-(k_\perp)} = \Bigg(\frac{\epsilon + \eta v_{ph}}{\epsilon - \eta v_{ph}}\Bigg)^2 .
\label{imbalance-chi=0} 
\end{equation}

{We  also get  $\rho^{3/2} = \dfrac{3\epsilon\sqrt{\beta_e}}{16\, k^4_\perp } \cosh^{-1} \frac{\psi}{2}$. For $k_\perp < k_i$, where we can assume $\psi = 0$ (in the absence of GCH injection at $k_0$), we can deduce from Eq. (\ref{eqphi-chi=0}, right) (using that the constant $C$ is large as a result of the small value of $\rho$ at $k_i$) that $\phi$ also tends to zero, as it is observed in the simulations of the diffusion model (Fig. \ref{inversecascade}, bottom). This explains the absence of propagation of the imbalance to large scales. }


\section{Transverse magnetic energy spectrum}\label{Bspectrum}
The transverse magnetic energy spectrum $E_{B_\perp}$, commonly measured in the SW, can be related, at least approximately, to the total energy spectrum. Writing ${\boldsymbol k}=(k_\perp, \theta, k_\|)$ in  cylindrical coordinates, the total energy can be expressed as ${\cal E}= \int E(k_\perp) dk_\perp$, with an energy spectrum given by $E(k_\perp) = (1/2) \int(s^2 |k_\perp L_e {\widehat A}_\||^2 + |k_\perp L_e \Lambda {\widehat \varphi}|^2) dk_\| k_\perp d\theta$, equivalent to Eq. (2.36) of  \cite{PS19}.
Here, ${\widehat \varphi}({\boldsymbol k})$ and ${\widehat A_\|}({\boldsymbol k})$ refer to the Fourier transforms  of the electrostatic and parallel magnetic potentials respectively,  $L_e = (1 + 2\delta^2 k_\perp^2/\beta_e)^{1/2}$ and $\Lambda$  defined in Eq. (2.17) of \citet{PS19}) is a function of $k_\perp$ which tends to $1$ as $k_\perp\to 0$ and is proportional to $1/k_\perp$ in the sub-ion range. The first term in the integral rewrites $L ^2_e E_{B_\perp}$, where $E_{B_\perp}= (s^2/2) \int |{\widehat B}_\perp({\boldsymbol k})|^2  dk_\| k_\perp d\theta$ corresponds to the magnetic energy spectrum, and  the second one reduces in the MHD regime to the kinetic energy spectrum. Their difference, referred to as the residual energy spectrum, is observed to remain small if initially zero in direct numerical simulations of the parent gyrofluid. In spite of the nonlinear interactions, the solution can indeed be viewed as a superposition of eigenmodes of both polarizations (which satisfy $s^2 |k_\perp L_e {\widehat A}_\||^2 = |k_\perp L_e \Lambda {\widehat \varphi}|^2$), and we are thus led to write $E_{B_\perp}(k_\perp) \approx  (1/2) L_e^{-2}E(k_\perp)$.

\bibliography{biblio}
\bibliographystyle{aasjournal}
\end{document}